\documentclass[letterpaper]{jpconf}
\usepackage{graphicx,amsmath}
\usepackage{graphics,graphicx,epsfig,color,amsmath}

\newcommand{\comment}[1]{\par\noindent {\em\small [#1]}}
\renewcommand{\comment}[1]{}

\newcommand{\labf}[1]{\label{fig:#1}}

\newcommand{\reff}[1]{\ref{fig:#1}}

\newcommand{\Figura}[1]{Figure~\reff{#1}}

\newcommand{\unit}[1]{\ensuremath{\rm\,#1}}

\newcommand{\mev}{\unit{MeV}}

\newcommand{\gev}{\unit{GeV}}
\newcommand{\gevc}{\unit{GeV\!/\!{\it c}}}

\newcommand{\tev}{\unit{TeV}}

\newcommand{\BAR}[1]{\overline{#1}}

\newcommand{\particle}[1]{{\ensuremath{\rm #1}}}

\newcommand{\pp}{{\em pp}}

\newcommand{\bb}{\particle{b\BAR{b}}}

\newcommand{\Bd}{\particle{B^0}}

\newcommand{\Bs}{\particle{B^0_s}}

\newcommand{\Jpsi}{\particle{J/\psi}}

\newcommand{\KS}{\particle{K^0_S}}

\newcommand{\Kst}{\particle{K^{*0}}}

\newcommand{\PT}{\ensuremath{\rm{P}_T}}

\begin{document}
\title{First LHCb Results from 2009 LHC Run}
\author{Kazuyoshi Carvalho Akiba\\
       { \em on behalf of the LHCb  Collaboration} \\
}

\address{Nikhef, Science Park 105, 1098 XG Amsterdam, The Netherlands}

\begin{abstract}
At the end of 2009, the Large Hadron Collider (LHC) provided a short run 
of \pp\ collisions  at  a centre-of-mass energy of $\sqrt{s} = 900 \gev$.
The LHCb Experiment collected its first collision data with the aim of finalizing the
commissioning of the detector and perform the spatial and time alignments.
This paper presents a collection of preliminary results of the LHCb detector obtained with the data acquired in  
this first LHC run. A brief outlook of the physics expected with 
the first data in 2010 at 7 \tev\ centre-of-mass energy is also presented. 

\end{abstract}

\section{Introduction}
The LHCb Experiment is dedicated to the precise reconstruction of B decays and the
study of CP violation in the $b$-quark sector. The angular acceptance  of the LHCb detector,  10 to 
300 mrad, 
was chosen because in 
high energy hadronic collisions  
the majority of \bb\ pairs are produced at low angles with respect 
to the collision axis. 
The LHCb detector is composed of several subsystems. 
Closest to the interaction region there is a   micro-strip
silicon detector dedicated to the reconstruction of primary and displaced vertices, called the Vertex Locator  (VELO). 
The is VELO built in two separate detector halves that retract to a fully open position,
when the LHC beams are being injected, and close around the interaction region when the
LHC beams become stable. The two most upstream stations in each VELO half constitute the Pile-Up System (PUS)
that is very similar to the VELO stations, but the PUS can read out ever bunch crossing (at 40 MHz)  and
hence are able to participate in the first level of the trigger. 
Hadron species can be distinguished by  
two Ring Imaging Cherenkov detectors (RICH) that  provide excellent particle identification  from 2 to 100 \gevc.
Calorimetry is provided by  a  fast  response Calorimeter system (CALO), composed of a Pre-Shower (PS), a Scintillating Pad Detector (SPD), the Hadron CALorimeter  (HCAL) and the 
Electromagnetic CALortimeter (ECAL), that participates in the first level (hardware) of trigger. The tracking system is built with two different
technologies, straw tube gas detectors in the lower occupancy regions (Outer Tracker, OT) and silicon micro strip detectors
in the higher occupancy ones (Silicon Tracker, ST) that together with  the magnet dipole gives very good momentum determination. 
The muon system composed of five stations to provide an excellent muon identification and also participates in the first trigger level.
The LHCb detector is fully described in \cite{jinst}.



\section{Trigger configurations}
\label{trigger}
In the 2009 pilot run the beam intensities and bunch filling were far lower than the nominal values for the LHC. A minimum 
bias trigger scheme was set up in order to record every collision. It required  at least one cluster above 
threshold (240 \mev) in the HCAL and at least three SPD cells with a hit.
In addition to the \pp\ trigger, a beam gas trigger based on the filling scheme and the information  
from either the  PUS or the HCAL+SPD decisions was also used. Since the LHCb Detector is designed as a single-arm spectrometer,
beam gas events coming from $-z$  \footnote{The $-z$ direction points from the Muon System towards the VELO, parallel to the beam axis.} 
can only be triggered by the PUS.
There were three distinct event types recorded:  beam1-gas with the beam coming from $+z$ direction, beam2-gas with 
the beam coming from the $-z$ direction,
and collision (or beam-beam). 

\section{General performance of the sub-detectors}
The first LHC run for physics started at the beginning of December 2009 and lasted for approximately two weeks. 
During that period, the LHCb detector  operated with all sub-detectors 
powered  and the  magnetic field at its nominal operational value.
The VELO collected data in a retracted position, 15 mm away from the nominal closed 
position\footnote{It means that the difference between the halves was equal to 30 mm.} due to the larger beam  
aperture at 450 $\gev$ energy per beam.  Around 300 thousand  \pp\ inelastic collisions events, 
corresponding to an integrated luminosity  of $L_{int} = 6.8 \pm 1.0~\rm{\mu b}^{-1}$, were 
acquired in the standard 2009 configuration which are useful for physics analyses. 


%
%


The statistical sample of 2009 was not large enough to align fully the mirrors of the RICH system 
nor to complete the time alignment of the photo-multipliers (the RICH uses HPDs \cite{jinst}). 
However the first data were suffcient to start these studies as a preparation for the 2010 run. 
In the run of 2009 it was also possible to measure the first kaon ``rings'' in the RICH system, 
and an example is shown in \Figura{kaons}.
\begin{figure}
\begin{center}
\includegraphics[width=0.30\textwidth]{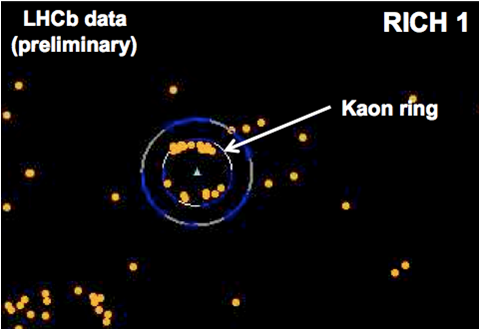}
\includegraphics[width=0.30\textwidth]{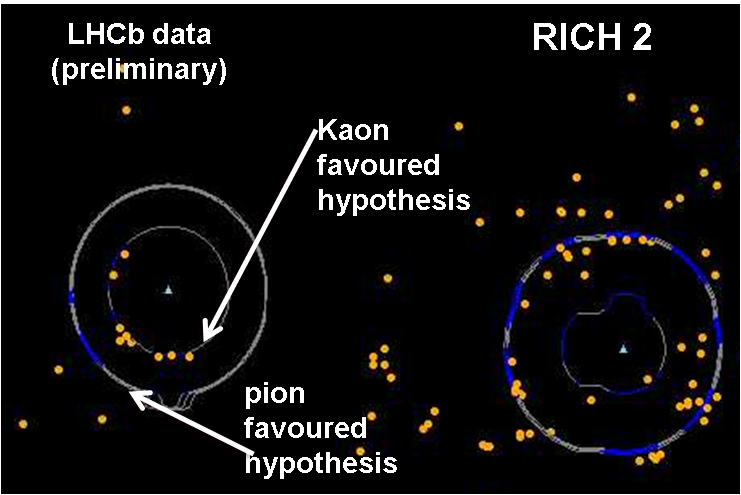}
\end{center}
\caption{Event display showing the reconstruction of rings in the RICH1 (left) and RICH2 (right). The smaller rings
are consistent with the Kaon hypothesis, as shown in the picture. }
\labf{kaons}
\end{figure}


The CALO proved to be reliable and efficient in the  run of 2009. One of the 
greatest accomplishments of the calorimeter group on LHCb was to time align the
four detector systems with a precision better than 2 ns, which is vital  due to its function in the
trigger. The CALO performs a few functions in LHCb besides participating in the trigger:
it provides the electron-photon separation and their measured position and energies; 
at the trigger level it selects events with a high \PT\ hadron arising from B decays. 
As can be seen in
\Figura{pizeroes}, already in the 2009 run the calorimeter calibration was good enough to enable
  a clear  $\pi^0$ signal to be isolated.
\begin{figure}
\begin{center}
\includegraphics[width=0.43\textwidth]{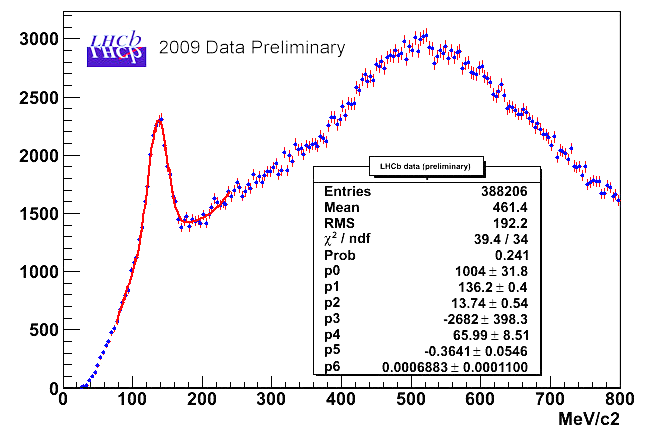}
\end{center}
\caption{The plot shows the mas reconstructed using photons in the ECAL. The $\pi^0$ peak 
is fitted to a Gaussian function on top of a polynomial background. The mean of the peak
is of $136 \pm 0.4$ \mev\  consistent with the the mass listed in  \cite{pdg}.  } 
\labf{pizeroes}
\end{figure}



Although not in the nominal closed position the VELO vertex reconstruction
performs reasonably well. The \Figura{beamgas} shows the 
beam profile in the $xz$ and $yz$ planes measured with protons
colliding to the residual gas inside the beam pipe.
\begin{figure}
\begin{center}
\includegraphics[width=0.30\textwidth]{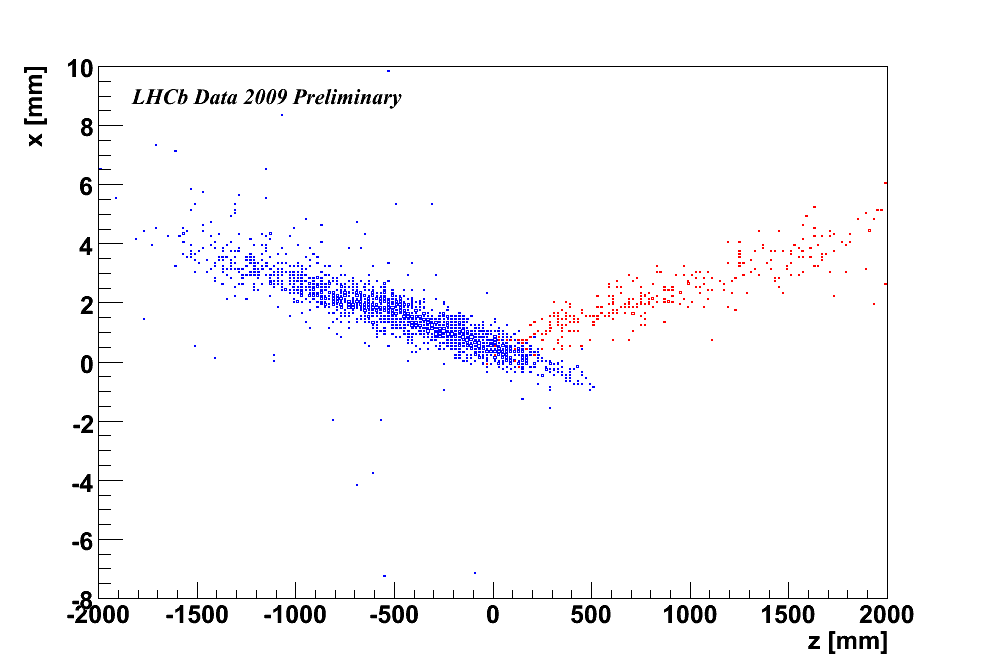}
\includegraphics[width=0.30\textwidth]{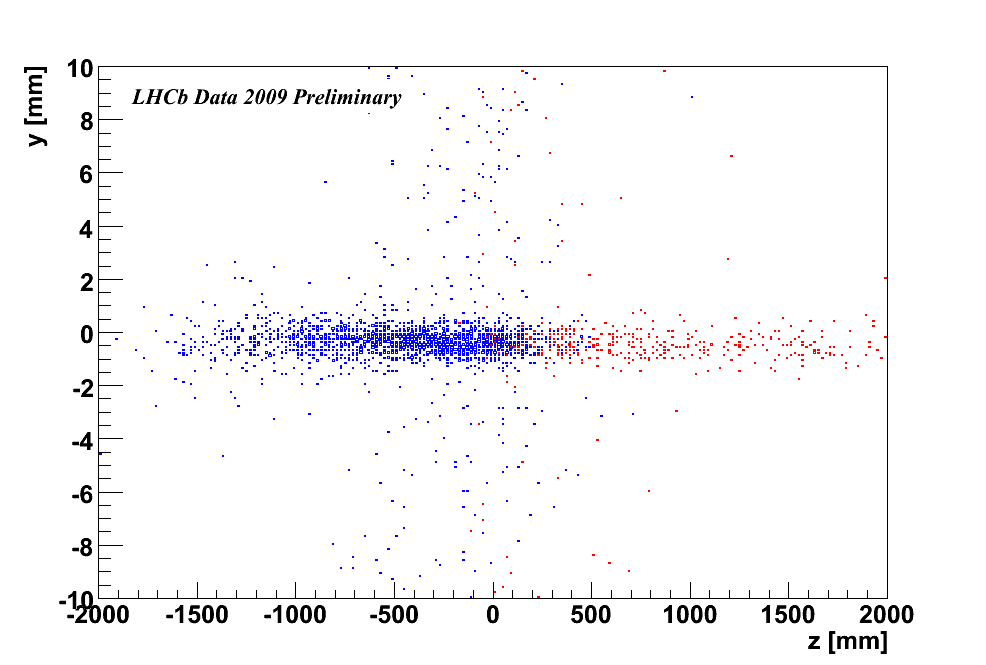}
\end{center}
\caption{The beam profiles   in the $xz$ (left) and $yz$ (right) planes (LHCb coordinates)
reconstructed using beam-gas events. The blue (red) points in the plot show the reconstructed 
vertex positions for beam1(2)-gas interactions.} 
\labf{beamgas}
\end{figure}

\section{V$^0$ Reconstruction}
The first data were used to reconstruct 
 V$^0$ particles
such as \KS\ and $\Lambda^0$ for the first time in the 2009 run. \Figura{ksandlambdas} 
shows the \KS\ and  $\Lambda^0$  mass peaks and compares the 
reconstruction of \KS\ with and without VELO tracks. The mass resolution will improve
with the alignment constants that are produced for the first time with the whole detector.
Since the conference LHCb has produced preliminary results
  on the differential cross-section of \KS\ production using these data \cite{KS}. 
\begin{figure}
\begin{center}
\includegraphics[width=0.32304\textwidth]{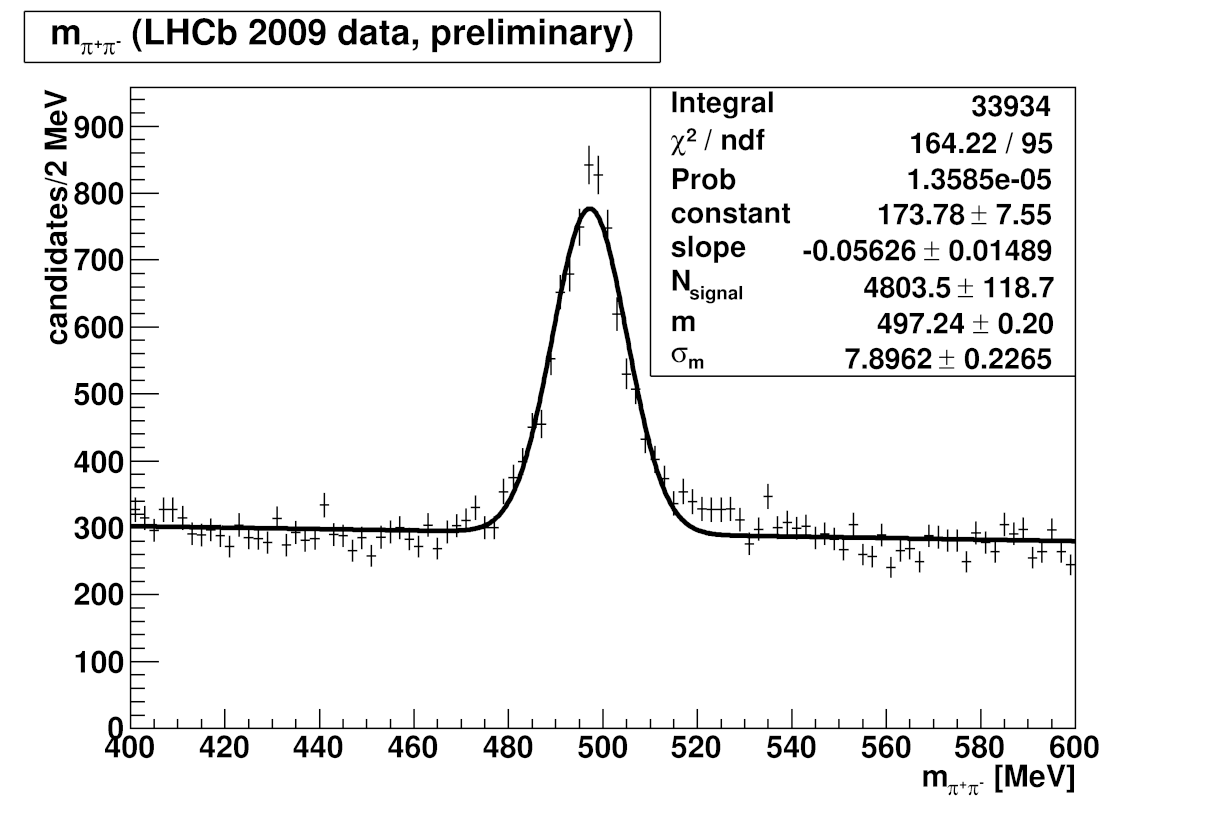}
\includegraphics[width=0.32304\textwidth]{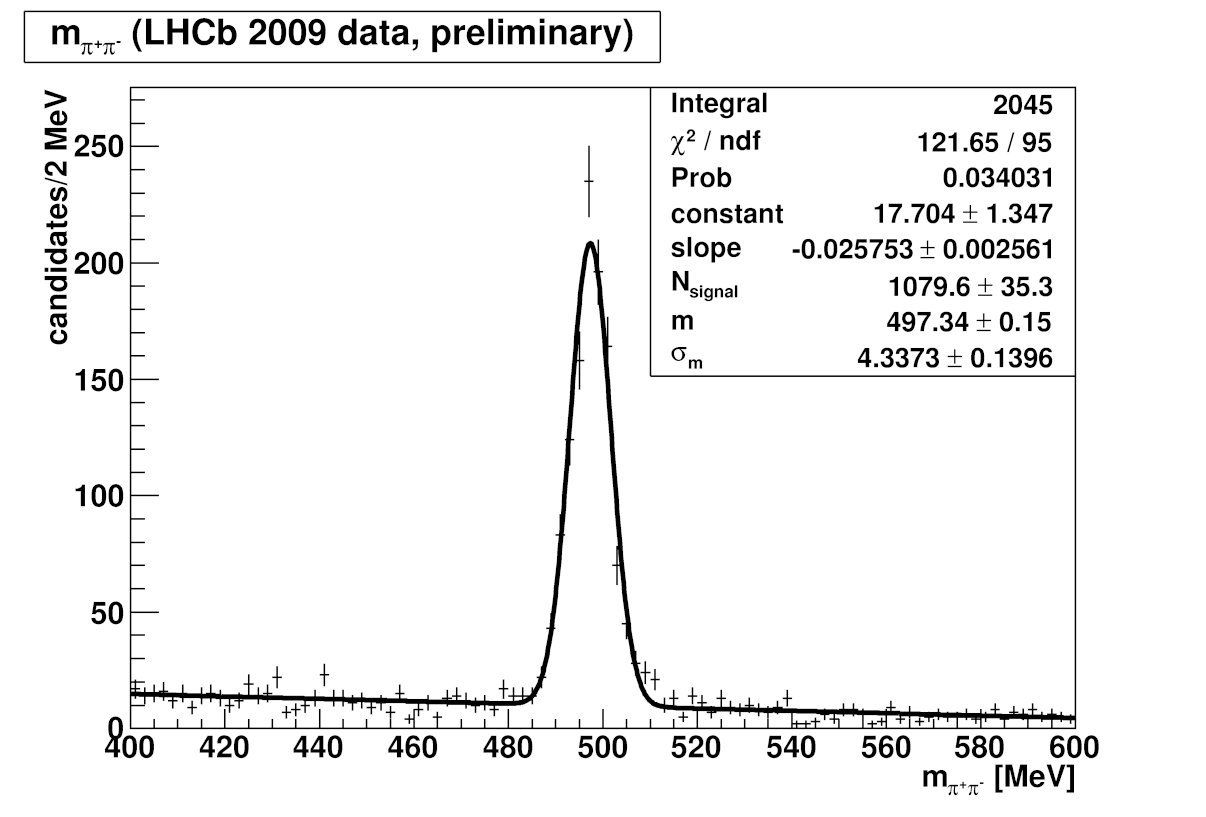}\\
\includegraphics[width=0.32304\textwidth]{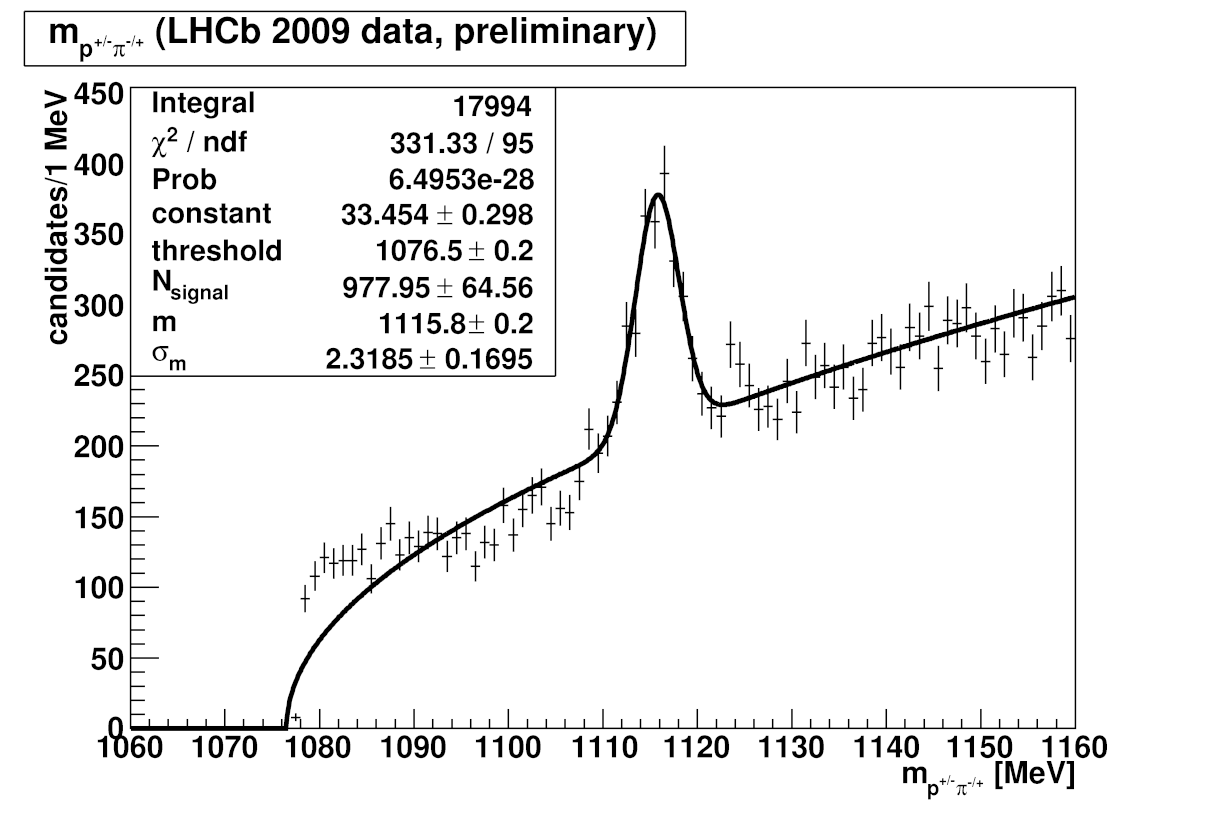}
\includegraphics[width=0.32304\textwidth]{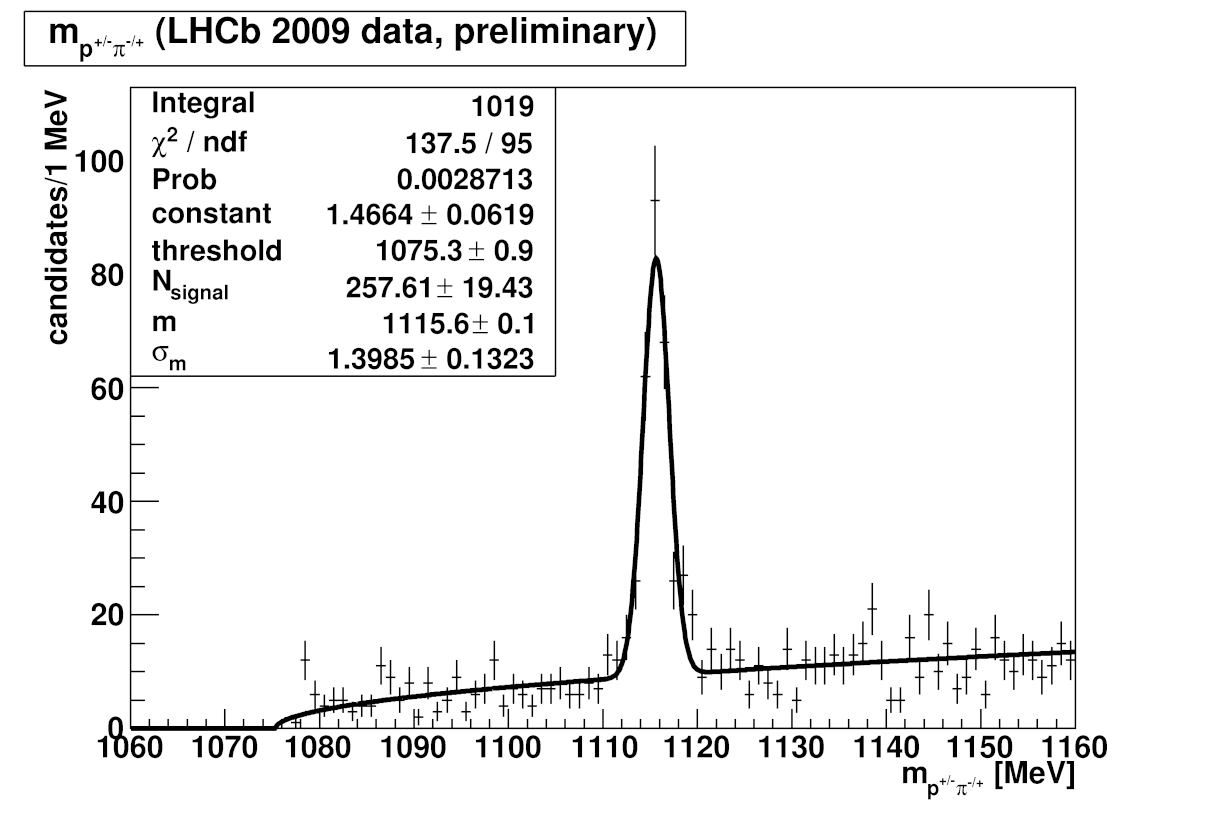}
\caption{The \KS\ (top) and  $\Lambda^0$ (bottom) mass reconstructed using all available tracks (left) and with tracks that have a VELO segment (right). 
The mass indicated in the plots is compatible with the best measurements listed in \cite{pdg}, noting that with the 
use of VELO tracks the mass resolution improves by more than a factor 2.  }
\end{center}
\labf{ksandlambdas}
\end{figure}
 
\section{Prospects for 2010-2011}
The  LHC plans to provide 1 $\rm{fb}^{-1}$ integrated luminosity 
at the centre-of-mass energy of $\sqrt{s} = 7 \tev$ in the run starting in 2010.
 For this amount of statistics the 
LHCb detector has an intensive physics programme \cite{roadmap}. Among many other measurements 
there are the following key goals:
precise measurement of $\phi_s$ with $\Bs\to\Jpsi\phi$ decay; perform a measurement of the 
forward-backward asymmetry $A_{FB}$ with the $\Bd\to\Kst\mu^+\mu^-$ channel;
make a precise measurement of the CKM phase $\gamma$ from tree dominated processes; probe for super-symmetry 
with the measurement of the branching ratio of the rare decay $\Bs\to\mu^+\mu^-$.
For the decay, $\Bs\to\mu^+\mu^-$, the Standard Model predicts
a very precise branching fraction: $3.55 \pm 0.33 \times 10^{-9}$ \cite{bsmumusmprediction}. Current searches in 
the Tevatron made by the CDF and D0 collaborations set already a few upper limits as reported in \cite{CDF}. 
Many new physics models  (like super symmetry) can generate higher branching ratios  
slightly below the $10^{-8}$ region  which the LHCb detector should be able to
measure a first evidence with approximately  1 $\rm{fb}^{-1}$, as shown in the plot of \Figura{bsmumu}. 
The precise measurement of this branching fraction can be made as more statistics are gathered over the following years.
\begin{figure}
\begin{center}
\includegraphics[width=0.38\textwidth]{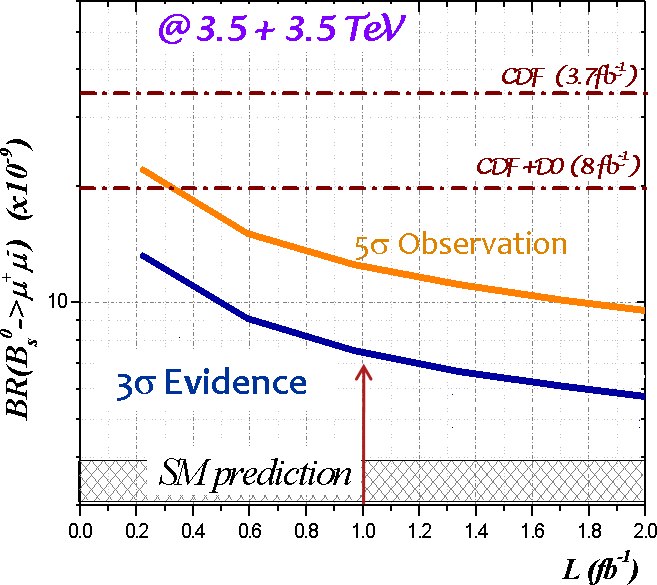}
\end{center}
\caption{Prospects for the branching ratio measurement of the $\Bs\to\mu^+\mu^-$ decay channel as a function of integrated luminosity. }
\labf{bsmumu}
\end{figure}
\section{Conclusions}
The LHCb detector had a very  fruitful first run in 2009. All the sub-systems have
proved to be ready to acquire data and coherently reconstruct events including decaying
particles. The first data in 2009 allowed  the  sub-detectors to perform their calibration
and (time) alignment adjustments, finalizing the LHCb commissioning. The 2010-2011 run is 
eagerly expected, in which the LHCb experiment will be able to show its first B-physics results.

\end{document}